\documentclass[a4paper,10pt]{article}
\usepackage{amsfonts, epsfig, amsmath, amssymb, color}

\newcommand{\I}[1]{\mathbb{I}\{#1\} }
\newcommand{\Ii}[2]{\mathbb{I}_{#1}(#2) }
\newcommand{\E}{\mathbf{E}}

\DeclareMathSymbol{\ophi}{\mathalpha}{letters}{"1E}

\renewcommand {\epsilon}{\varepsilon}

\renewcommand{\phi}{\varphi}

\newcommand{\be}{\begin{equation}}
\newcommand{\ee}{\end{equation}}
\newcommand{\ben}{\begin{equation*}}
\newcommand{\een}{\end{equation*}}

\newcommand{\ba}{\begin{equation}\begin{aligned}}
\newcommand{\ea}{\end{aligned}\end{equation}}
\newcommand{\ban}{\begin{equation*}\begin{aligned}}
\newcommand{\ean}{\end{aligned}\end{equation*}}

\DeclareMathOperator{\sgn}{sgn}

\allowdisplaybreaks[4]

\begin{document}
\title{L\'evy Flights, Non-local Search and Simulated Annealing}

\author{Ilya Pavlyukevich\footnote{Correspondence should be addressed to pavljuke@mathematik-hu-berlin.de}
}

\date{22 January 2007}


\maketitle
\begin{abstract}
We solve a problem of non-convex stochastic optimisation with help of simulated annealing of 
L\'evy flights of a variable stability index.
The search of the ground state of an unknown potential is non-local 
due to big jumps of the Levy flights process. 
The convergence to the ground state is fast due to a
polynomial decrease rate of the temperature. 

\end{abstract}


\numberwithin{equation}{section}

\section{Introduction}

Let $U$ be a potential function in $\mathbb{R}^d$ having several local minima and increasing fast at infinity. We look for a global minimum of $U$.
Classical continuous-time simulated annealing (Boltzmann machine) 
(see \cite{VanderbildL-84,Gidas-85,GemanH-86,GelfandM-93}) 
consists in running a diffusion process
\ba
\hat{Z}_{0,z}(t)&=z-\int_0^t\nabla U(\hat{Z}_{0,z}(u))\,du+\int_0^t\hat{\sigma}(u) dW(u),\\
\hat{\sigma}(t)&=\left(\frac{\theta}{\ln(\lambda+u)}\right)^{\!1/2},
\ea
where $W$ is a standard Brownian motion, $\theta>0$ denotes the cooling rate and $\lambda>1$ parametrises the initial temperature,
which equals $\sqrt{\theta/\ln(\lambda)}$ at time $t=0$.
It is known that there is a critical value $\hat{\theta}$ such that 
the diffusion $\hat{Z}(t)$ converges in distribution
to the \textit{global} minimum of $U$ if $\theta>\hat{\theta}$, and 
the convergence fails otherwise.
Moreover, the critical value $\hat{\theta}$ is the logarithmic rate of the principal non-zero eigenvalue $\lambda_1(\sigma)$ 
of a time homogeneous diffusion generator 
$A_\sigma f=\frac{\sigma^2}{2}\Delta f-\langle\nabla U, f\rangle$, i.e.\
\be
\label{eq:hattheta}
\hat{\theta}=-\lim_{\sigma\to 0}\sigma^2\ln|\lambda_1(\sigma)|.
\ee
The value of $\hat{\theta}$ can be calculated explicitly, if one knows the heights
of potential barriers between different wells of $U$ (see \cite{Wentzell-72,Wentzell-72a} and \cite{KolokoltsovM-96} for precise results).
Rigorous results on optimal cooling rate in simulated annealing algorithms can
be found 
in \cite{Hajek-88,ChiangHS-87,HolleyKS-89,HwangS-90}.

In order to accelerate the search, Szu and Hartley in \cite{SzuH-87} suggested the 
so-called \textit{fast simulated annealing} (Cauchy machine), which is a combination of a 
classical Metropolis algorithm introduced in \cite{MetropolisRRT-53} and a concept of non-local search 
due to the heavy-tail Cauchy visiting distribution.
The authors claimed that in the Cauchy machine the temperature can be chosen decreasing as a power of time, namely $\sigma(t)\sim t^{-1}$,
and applied the algorithm in image processing, see \cite{Szu-93}.

Motivated by \cite{SzuH-87}, in our papers \cite{Pavlyukevich-06a,Pavlyukevich-06b} we considered a 
continuous-time counterpart of the process $\hat{Z}$
driven by \textit{L\'evy flights} of stability index $\alpha\in(0,2)$, and temperature $\sigma(t)\sim t^{-\theta}$, $\theta>0$. We 
discovered that such a jump process never settles in the neighbourhood of a global minimum, but can be used to reveal a spatial  
structure of the potential $U$. The dynamics of L\'evy flights with constant 
small noise was studied in our previous papers 
\cite{ImkellerP-06,ImkellerP-06a,ImkellerP-06b}.

In the present paper we solve the problem of global optimisation with help of \textit{state-dependent} 
L\'evy flights in a multi-well potential $U$. We show, that
in certain annealing regimes, the global minimum is localised always, as in the classical Gaussian case.
For simplicity, we restrict our theoretical argument in Sections~\ref{s:homogen}--\ref{s:search} to one-dimensional potentials. However, it will be clear from the
presentation, that the algorithm also works in a multi-dimensional setting. In our numerical example in 
Section~\ref{s:numeric} we consider a two-dimensional potential with five wells. 

\section{Results on the cooled down L\'evy flights\label{s:homogen}}

In \cite{Pavlyukevich-06a,Pavlyukevich-06b} we considered a one-dimensional L\'evy flights process in an external potential $U$ determined by the equation
\be
\label{eq:Z}
Z^{(\alpha)}_{0,z}(t)=z-\int_0^t U'(Z^{(\alpha)}_{0,z}(u-))\,du+\int_0^t\frac{dL^{(\alpha)}(u)}{(\lambda+u)^\theta}.
\ee
We understand a L\'evy flights process $L^{(\alpha)}$ as
a symmetric stable L\'evy process with 
stability index $\alpha\in(0,2)$, whose
marginal distributions have the Fourier transform
\be 
\label{eq:L}
\mathbf{E} e^{i\omega L^{(\alpha)}(t)} = e^{-c(\alpha)t|\omega|^\alpha}
=\exp\left(t\int_{\mathbb{R}\backslash\{0\}}\left[e^{i\omega y}-1-i\omega y\Ii{D}{y} \right]\frac{dy}{|y|^{1+\alpha}} \right),
\ee
where $\Ii{D}{y}$ is the indicator function of the unit disk $D=\{y: |y|\leq 1\}$, and $c(\alpha)= 2|\cos\left( \frac{\pi\alpha}{2}\right)\Gamma(-\alpha)|$.
We choose such a parametrisation in order to have a simple form of a L\'evy measure $\nu(dy)=|y|^{-1-\alpha}dy$.

The measure $\nu$ is also called a jump measure of the process L\'evy $L^{(\alpha)}$. 
It controls the intensity and sizes of its jumps. Indeed, let 
$\Delta L^{(\alpha)}(t)=L^{(\alpha)}(t)-L^{(\alpha)}(t-)$ denote the jump size
of $L^{(\alpha)}$ at time instance $t>0$. 
Then the number of jumps on the time interval $(0,t]$ with values in a set 
$J\subseteq \mathbb{R}$ is a Poissonian random variable with the mean
$t\nu(J)$ (which can be possibly zero or infinite).

The process $L^{(\alpha)}$ is a Markov process with a non-local generator
\be
Af(x)=\int_{\mathbb{R}\backslash\{0\}}\left[f(x+y)-f(x)-yf'(x)\Ii{D}{y} \right]\frac{dy}{|y|^{1+\alpha}},
\ee
which is also referred to as a fractional Laplacian, $A=\Delta^{\alpha/2}$.

We direct reader's attention to the books 
\cite{Sato-99,Applebaum-04,Protter-04} on a rigorous mathematical theory of 
L\'evy processes and stochastic differential equations. 
Physical results on the subject can be found in 
\cite{MetzlerK-04,ChechkinGKM-04,BrockmannS-02}.

We assume that the potential $U$ has $n$ local minima $m_i$ and $n-1$ local maxima $s_i$ such that 
$-\infty=s_0<m_1<s_1<\cdots <m_n<s_{n+1}=+\infty$. The extrema are non-degenerate, i.e.\ $U''(m_i)>0$ and $U''(s_i)<0$. Moreover
we demand that $|U'(x)|>|x|^{1+c}$ as $|x|\to+\infty$ for some positive $c$.

In the small temperature limit, i.e.\ when $\lambda\to+\infty$ or $t\to+\infty$,
the process
$Z^{(\alpha)}$ can be seen as a random perturbation of a deterministic dynamical system
\be
X^0_x(t)=x-\int_0^t U'(X^0_x(s))\,ds.
\ee
We denote $\Omega_i=(s_{i-1},s_i)$, $1\leq i\leq n$, the domains of attraction of the stable points $m_i$.

The positive parameter $\theta$ is called the cooling rate, and $\lambda>0$ determines the initial temperature of the system
which equals to $\lambda^{-\theta}$ at $t=0$.

Equation \eqref{eq:Z} describes a non-linear dynamics of a L\'evy particle, whose temperature is being decreased at a 
polynomial rate as $t\to\infty$. 

In \cite{Pavlyukevich-06a,Pavlyukevich-06b}, we discovered two cooling 
regimes --- slow cooling $\theta<1/\alpha$ and fast cooling $\theta>1/\alpha$ --- in
which the transitions of a particle between the wells of $U$ have different asymptotic properties.

Let $\Delta>0$ be a small number, and let $B_i=\{y:|y-m_i|\leq \Delta\}$ denote a $\Delta$-neighbourhood of a local minimum $m_i$.
Consider transition times 
\be
T_{s,z}^{i}=\inf\{u\geq s:Z_{s,z}^{(\alpha)}(u)\in \cup_{j\neq i}B_j\}
\ee
between different neighbourhoods $B_i$ 
and the corresponding transition probabilities
 $\mathbf{P}_{s,z}(Z^{(\alpha)}(T^{i,\lambda})\in B_j)$, $i\neq j$. Then for $\theta<1/\alpha$ and 
$z\in B_i$ we have 
\begin{align}
\label{eq:tau}
&\frac{\mathbf{E}_{0,z} T^{i}}{\lambda^{\alpha\theta}}\to [q_i^{(\alpha)}]^{-1},
\quad \lambda\to+\infty,\\
\label{eq:p}
&\mathbf{P}_{0,z}(Z^{(\alpha)}(T^{i})\in B_j)
\to q_{ij}^{(\alpha)}[q_i^{(\alpha)}]^{-1},\quad i\neq j,
\end{align}
where
\ba
\label{eq:q}
q_{ij}^{(\alpha)}
&=\int_{\Omega_j}\frac{dy}{|m_i-y|^{1+\alpha}}
=\frac{1}{\alpha}
\left|\frac{1}{|s_{j-1}-m_i|^\alpha}-\frac{1}{|s_j-m_i|^\alpha}\right|,\quad i\neq j,\\
q_i^{(\alpha)}&=
\int_{\mathbb{R}\backslash\Omega_i}\frac{dy}{|m_i-y|^{1+\alpha}}
=\sum_{j\neq i}q_{ij}^{(\alpha)}=\frac{1}{\alpha}\left( \frac{1}{|s_{i-1}-m_i|^\alpha}
+\frac{1}{|s_i-m_i|^\alpha}\right).
\ea
We have also shown that in the limit $t\to+\infty$, $Z^{(\alpha)}_{0,z}(t)$ has  a distribution 
\be
\pi^{(\alpha)}(dy)=\sum_{i=1}^n\pi^{(\alpha)}_i\delta_{m_i}(dy)
\ee
where the vector
$\pi^{(\alpha)}=(\pi^{(\alpha)}_1,\dots,\pi^{(\alpha)}_n)^T$, 
solves the equation $Q^T\pi^{(\alpha)}=0$, 
$Q=(q_{ij}^{(\alpha)})_{i,j=1}^n$, $q_{ii}^{(\alpha)}=-q_i^{(\alpha)}$.
It is clearly seen, that all $\pi^{(\alpha)}_i>0$, and $Z^{(\alpha)}$ does not settle down near the global minimum of $U$. However, 
the values $\pi^{(\alpha)}_i$, which can be estimated from the Monte Carlo simulations, reveal the spatial structure of $U$, e.g.\
the sizes of the domains $\Omega_i$.

If the cooling rate $\theta$ is above the threshold $1/\alpha$, the L\'evy particle $Z^{(\alpha)}$ gets trapped in 
one of the wells and thus the convergence fails. Consider the first exit time 
from the $i$-th well
\be
S^{i}_{s,z}=\inf\{u\geq s: Z^{(\alpha)}_{s,z}\notin \Omega_i \}.
\ee
Then, for $z\in B_i$, $1\leq i\leq n$,  
\be
\label{eq:S}
\mathbf{P}_{0,z}(S^{i}<\infty)
=\mathcal{O}\left(\frac{1}{\lambda^{\alpha\theta-1}}\right),\quad\lambda\to\infty,
\ee
and consequently, $\mathbf{E}_{0,z}S^{i}=\infty$.

\section{L\'evy flights with variable stability index (stable-like processes)}

In order to take into account the energy geometry of the potential, we have to make the L\'evy flights process
depend on its current position. Thus instead of L\'evy flights 
$L^{(\alpha)}$ defined in \eqref{eq:L}, we consider now the so-called
\textit{stable-like process} $H=(H(t))_{t\geq 0}$, which is a Markov process defined by the non-local generator
\be
Bf(x)=\int_{\mathbb{R}\backslash\{0\}}\left[f(x+y)-f(x)-yf'(x)\Ii{D}{y} \right]\frac{dy}{|y|^{1+\alpha(x)}},
\ee
with a function $\alpha(x)$ taking values in the interval $(0,2)$. Sometimes, the notation $B=\Delta^{\alpha(\cdot)/2}$
is used. 
The main difference between $L^{(\alpha)}$ and 
$H$ consists in a dependence of a stable-like jump measure $\nu_x(dy)=|y|^{-1-\alpha(x)}$
on the spatial coordinate $x$.
Thus, if $H(t_0)=x_0$, the instant jump distribution of $H$ at time $t_0$ is governed by a 
stable measure $\nu_{x_0}(dy)$. 

Of course, the dynamics of $H$ is completely determined by a variable stability index $\alpha(x)$.
For example, if $\alpha(x)=\alpha_0\in(0,2)$, then $H$ is just a usual L\'evy flights process
of index $\alpha_0$. From now on, we assume that $\alpha(x)$ takes values strictly between 
$0$ and $2$, i.e.\ $0<a\leq \alpha(x)\leq A<2$, to exclude
degeneration of the jump measure.

We are going to study the dynamics of a stochastic differential equation with the driving 
process $H$, namely
\be
\label{eq:Y}
Y_{0,y}(t)=y-\int_0^t U'(Y_{0,y}(u-))\,du
+\int_0^t\frac{dH(Y_{0,y}(u-),u)}{(\lambda+u)^\theta}.
\ee
For a better understanding of the process $Y$, it is instructive to consider a discrete time analogue of \eqref{eq:Y} 
given by the recurrent formula
\ba
\label{eq:y}
y_{k}&=y_{k-1}-U'(y_{k-1})h+\frac{\xi_{k}^h(y_{k-1})}{(\lambda+(k-1)h)^\theta},\quad k\geq 1.
\ea
The discrete time dynamical system \eqref{eq:y} is obtained from the Euler approximation of \eqref{eq:Y} with the time step $h$, 
and can be used for simulations. (However, one should be careful when $U'$ is not globally Lipschits.) The random 
input is determined by the random variables $\xi_k(y)$ such that
\be
\xi_k^h(y)\stackrel{d}{=}L^{(\alpha(y))}(h)\stackrel{d}{=}h^{1/\alpha(y)}L^{(\alpha(y))}(1)
\ee
where $L^{(\alpha(y))}(1)$ has a standard symmetric $\alpha(y)$-stable distribution 
with the Fourier transform \eqref{eq:L}.

\section{One-well dynamics. Transitions and trapping}

The dynamics $Y$ of the L\'evy flights with variable stability index in the 
force field $U'$ is a result of an interplay of two 
independent effects. First, for small temperatures, 
i.e.\ when $\lambda\to+\infty$ or $t\to+\infty$, $Y$ is close to the underlying deterministic
trajectory $X^0$. Starting from any point of $\Omega_i$, it follows 
$X^0$ with the same
initial point and with high probability reaches a small neighbourhood of $m_i$ in relatively short time.
On the other hand, $Y$ tries to deviate from $X^0$ making jumps controlled by 
the jump
measure $\nu_x(dy)$. Finally, if $Y$ is in the well $\Omega_i$, it spends most of the time  
in a neighbourhood of $m_i$ and thus has jumps approximately governed by the stable 
jump measure $\nu_{m_i}(dy)=|y|^{-1-\alpha(m_i)}dy$.

Thus, the exit time and the exit probability from the well $\Omega_i$ 
of the process $Y$ are approximately the same as for the process $Z^{(\alpha(m_i))}$.
This resemblance becomes exact if we consider a piece-wise constant stability index $\alpha(x)$, 
$\alpha(x)=\sum_{i=1}^n\alpha_i\I{x\in \Omega_i}$, $0<\alpha_i<2$. 
With this choice of $\alpha(x)$, the process $Y$ is just driven by the equation
\eqref{eq:Z} until it exits the well.
(We omit a discussion on the behaviour of the process in the small neighbourhoods of the saddle points.)

Let us introduce the following transition and exit times for the process $Y$: 
\begin{align}
\label{eq:1}
\tau_{s,z}^{i}&=\inf\{u\geq s:Y_{s,z}(u)\in \cup_{j\neq i}B_j\},\\
\sigma^{i}_{s,z}&=\inf\{u\geq s: Y_{s,z}\notin \Omega_i \}.
\end{align}
It follows from \eqref{eq:tau} and \eqref{eq:p}, that if $y\in B_i$ and $\alpha(m_i)\theta<1$, then the following relations hold as $\lambda\to+\infty$:
\begin{align}
\label{eq:2}
&\frac{\mathbf{E}_{0,y}\tau^{i}}{\lambda^{\alpha(m_i)\theta}}\to [q_i^{(\alpha(m_i))}]^{-1},\\
\label{eq:3}
&\mathbf{P}_{0,z}(Y(T^{i})\in B_j)
\to q_{ij}^{(\alpha(m_i))} [q_i^{(\alpha(m_i))}]^{-1},\quad i\neq j.
\end{align}
On the other hand, if $\alpha(m_i)\theta>1$, the L\'evy particle gets trapped in the well due to \eqref{eq:S},
i.e.\
\be
\label{eq:4}
\mathbf{P}_{0,z}(\sigma^{i}<\infty)
=\mathcal{O}\left(\frac{1}{\lambda^{\alpha(m_i)\theta-1}}\right),\quad\lambda\to+\infty,
\ee
and consequently, $\mathbf{E}_{0,z}\sigma^{i,\lambda}=\infty$.

Relations \eqref{eq:2}, \eqref{eq:3} and \eqref{eq:4} are crucial for our analysis.

\section{Non-local random search and simulated annealing\label{s:search}}

\subsection{Getting trapped in an assigned well}

We demonstrate now how to drive the L\'evy particle $Y$ 
to an assigned well $\Omega_i$, if the
approximate location of its minimum $m_i$ is known. 

Indeed, the function $\alpha(x)$ given, the global dynamics of $Y$ is determined by 
the values $\alpha(m_i)$, $1\leq i\leq n$, and the cooling rate $\theta$. Moreover,
for our analysis we can freely choose both $\alpha(x)$ and $\theta$.

Let $\alpha(x)$ be smooth and attain its unique global maximum at $m_i$. Then we have 
\be
\alpha(m_i)>\max_{j\neq i}\alpha(m_j).
\ee
For instance, one can take $\alpha(x)=(A-a)/(1+(x-m_i)^2)+a$ for some $0<a<A<2$.
Then we can choose $\theta>0$, such that
\be
\alpha(m_i)\theta>1,\quad\mbox{whereas}\quad \alpha(m_j)\theta<1
\,\,\text{ for }\,\, j\neq i.
\ee 
With this choice of parameters, as $t\to+\infty$, the particle leaves any well 
$\Omega_j$, $j\neq i$, in a finite time according to \eqref{eq:2}.
Moreover, since all transition probabilities in \eqref{eq:3} 
are strictly positive, 
the probability to enter the well $\Omega_i$ after a finite number 
of transitions between the wells $\Omega_j$, $j\neq i$, equals $1$. 
Finally, upon entering $\Omega_i$, the particle gets trapped there due to \eqref{eq:4}. 

\subsection{Looking for the global minimum\label{ss:globmin}}

Let $M$ be the (unique) unknown global minimum of the potential $U$. To make a L\'evy particle
settle near it, we have to determine the appropriate variable stability index 
$\alpha(x)$ and the cooling rate $\theta$ such that
\be
\label{eq:al}
\alpha(M)\theta>1,\quad\mbox{whereas}\quad \alpha(m_i)\theta<1, \,\, m_i\neq M.
\ee 
Let $\phi(u)$ be an \textit{arbitrary} smooth monotone \textit{decreasing} function on $\mathbb{R}$, $0<a<\phi(u)<A<2$.
Then we set 
\be
\alpha(y)=\phi(U(y)).
\ee
It is clear that
\be
\alpha(M)>\alpha(m_i)\quad \mbox{for all }\,\, m_i\neq M,
\ee 
and we choose $\theta$ to satisfy
relations \eqref{eq:al}. The trapping of the particle near the global minimum $M$
follows from the argument of the preceding section. 

\subsection{The local minimum with maximal energy}

Analogously, we can determine the coordinate of the (unique) `highest' local minimum $m$, i.e.\
such that 
\be
U(m)=\max_{1\leq i\leq n} U(m_i).
\ee
In this case, we should take $\alpha(y)=\psi(U(y))$ with an \textit{arbitrary} smooth monotone \textit{increasing}  function $\psi$, $0<a<\phi(y)<A<2$, which leads to the inequality
\be
\alpha(m)>\alpha(m_i)\quad \mbox{for all }\,\, m_i\neq m,
\ee 
and the arguments of the previous sections justify the success of the search.

\subsection{A local minimum with certain energy \label{ss:energy}}

Finally, we can perform a search not for the global minimum of $U$ 
but for a local minimum
$m_E$ satisfying the condition $U(m_E)\leq E$ for some energy level $E$.
In such a case we take a piece-wise constant stability index
\be
\alpha(y)=
\begin{cases}
A,\quad U(y)\leq E,\\
a,\quad U(y)\geq E, \quad 0<a<A<2,
\end{cases}
\ee
and a cooling rate $\theta$ satisfying conditions $A\theta>1$ and $a\theta<1$. 
Consequently, if for some $i$, the minimum $U(m_i)$ lies below the threshold $E$, then near this minimum $Y$ behaves like
a jump-diffusion $Z^{(A)}$ driven by a L\'evy flights process $L^{(A)}$, 
and thus the L\'evy particle gets trapped due to \eqref{eq:4}. 
If the well minimum lies above the level $E$, a L\'evy particle behaves like
a process $Z^{(a)}$ and leaves this well in finite time due to relations
\eqref{eq:2} and \eqref{eq:3}.

If there are several wells with minima below $E$, the L\'evy particle settles down near one of them.  

Running the search for decreasing
energies $E_1>E_2>...$, we can also estimate the ground energy level $E^*=U(M)$ and to determine the global minimum
$M$.
Analogously, one can determine a local minimum $m_E$ satisfying conditions $U(m_E)\geq E$ or $E_1\leq U(m_E)\leq E_2$.

We emphasise that the search algorithm described in this section requires no particular information about the potential $U$: to determine the local minimum $m_E$
we use three numbers $a$, $A$ and $\theta$, the energy level $E$, and values
of $U'(x)$.


\section{Numerical example \label{s:numeric}}

In the preceding sections we gave a theoretic justification of the search algorithm in dimension one. 
It is clear that a $d$-dimensional case, $d>1$, does not differ much. It 
suffices 
to consider isotropic (spherically symmetric) $d$-dimensional 
L\'evy flights with a jumping measure 
$\nu_\mathbf{x}(d\mathbf{y})=|\mathbf{y}|^{-d-\alpha(\mathbf{x})}d\mathbf{y}$, and 
to calculate new values of
transition probabilities in \eqref{eq:q} 
writing $d$ instead of $1$ in the integrals.

Similarly to \eqref{eq:y}, we generate a $d$-dimensional Markov chain  
\ba
\label{eq:yd}
\mathbf{y}_{k}&=\mathbf{y}_{k-1}-\nabla U(\mathbf{y}_{k-1})h
+\frac{\xi_{k}^h(\mathbf{y}_{k-1})}{(\lambda+(k-1)h)^\theta},\quad 1\leq k\leq n,
\ea
with some initial value $\mathbf{y}_0$.
Isotropic $\alpha$-stable random vectors $\xi_k^h(\mathbf{y})$ are obtained as marginals of
a subordinated standard $d$-dimensional Brownian motion (e.g., see Examples~24.12 and 30.6 in \cite{Sato-99}).
More precisely, let $S$ be an $\alpha/2$-stable strictly positive random variable with the Fourier transform
\ba
\E \exp(i \omega S)
=\exp\left( -|\omega|^{\alpha/2}(1-i\sgn(\omega)\tan(\frac{\pi\alpha}{4}))\right),\quad \omega\in\mathbb{R},
\ea
and the Laplace transform
\ba
\E \exp(-u S)=\exp\left( -  \frac{u^{\alpha/2}}{\cos(\frac{\pi\alpha}{4})}\right),\quad u\geq 0.
\ea
Let $\mathbf{W}$ be a standard Gaussian vector independent of $S$ with a characteristic function 
\ba
&\E \exp\left( i\langle\mathbf{\omega},\mathbf{W}\rangle\right) =\exp\left(-\frac{|\omega|^2}{2} \right),
\quad\omega\in\mathbb{R}^d.
\ea
Then one obtains a standard isotropic $d$-dimensional $\alpha$-stable random vector $\mathbf{L}^{(\alpha)}$ 
with the Fourier transform  
\ba
\label{eq:L2}
\mathbf{E} \exp\left( i \langle\omega, \mathbf{L}^{(\alpha)}\rangle\right) 
&=\exp\left(\int_{\mathbb{R}^d\backslash\{0\}}\left[e^{i(\omega,\mathbf{y})}-1
-i(\omega, \mathbf{y})\Ii{D}{\mathbf{y}} \right]
\frac{d\mathbf{y}}{|\mathbf{y}|^{d+\alpha}} \right)\\
&=\exp\left(- \frac{\pi^{d/2}}{2^\alpha}\frac{|\Gamma(-\frac{\alpha}{2})|}{\Gamma(\frac{d+\alpha}{2})}   |\omega|^\alpha     \right),
\quad \omega\in\mathbb{R}^d,
\ea
as
\be
\mathbf{L}^{(\alpha)}
\stackrel{d}{=}c(\alpha,d) \sqrt{S}\cdot\mathbf{W},\quad c(\alpha,d)
=\frac{1}{\sqrt{2}}\left[ \pi^{d/2} \cos\left(\frac{\pi\alpha}{4}\right)\frac{|\Gamma(-\frac{\alpha}{2})|}{\Gamma(\frac{d+\alpha}{2})} \right]^{1/\alpha} 
\ee
Finally the random increments in \eqref{eq:yd} can be calculated as
\be
\xi_k^h(\mathbf{y})=h^{1/\alpha(\mathbf{y})}\mathbf{L}^{(\alpha(\mathbf{y}))}.
\ee

\subsection{Five-well potential in $\mathbb{R}^2$}

To illustrate the efficiency of the method, we consider a two-dimensional potential function $U$ 
given by the formula
\ba
U(\mathbf{y})&=\Bigg[1- \frac{1}{1+0.05(y_1^2+(y_2-10)^2)}
-\frac{1}{1+0.05((y_1-10)^2+y_2^2)}
-\frac{1.5}{1+0.03((y_1+10)^2+y_2^2)}\\
&-\frac{2}{1+0.05((y_1-5)^2+(y_2+10)^2)}
-\frac{1}{1+0.1((y_1+5)^2+(y_2+10)^2)}\Bigg] (1+0.0001 (y_1^2+y_2^2)^{1.2}).
\ea
The function $U$ has five local minima $\mathbf{m}_i$, $i=1,\dots,5$, with the following coordinates and energy values:
\ba
\mathbf{m}_1&\approx (-9.73,-0.11),&&& U(\mathbf{m}_1)\approx -0.85,\\
\mathbf{m}_2&\approx (-0.09,\hphantom{-}9.64),&&& U(\mathbf{m}_2)\approx -0.44,\\
\mathbf{m}_3&\approx (\hphantom{-}9.59,-0.37),&&& U(\mathbf{m}_3)\approx -0.54,\\
\mathbf{m}_4&\approx (\hphantom{-}4.92,-9.89),&&& U(\mathbf{m}_4)\approx -1.46,\\
\mathbf{m}_5&\approx (-4.79,-9.79),&&& U(\mathbf{m}_5)\approx -0.79.
\ea 
We perform a search for a local minimum $\mathbf{m}_E$ having the energy less than $E=-1$, i.e.\ such that 
$U(\mathbf{m}_E)\leq -1$. 
In our example, $\mathbf{m}_E=\mathbf{m}_4$. According to Section~\ref{ss:energy}, we choose a piece-wise constant 
stability index 
\be
\alpha(\mathbf{y})=\begin{cases}
           1.8, \text{ if } U(\mathbf{y})< -1,\\
           1.1, \text{ if } U(\mathbf{y})\geq -1,
          \end{cases}
\ee 
and $\theta=0.75$.

\begin{figure}
\begin{center}
\epsfig{figure=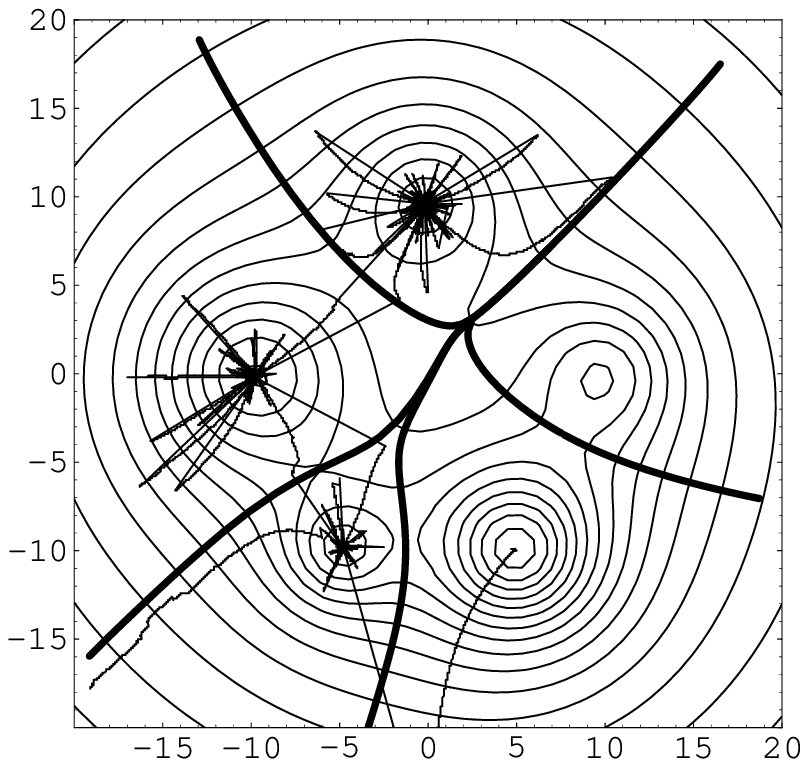, height=.25\textheight}
\hspace{.05\textheight}\epsfig{figure=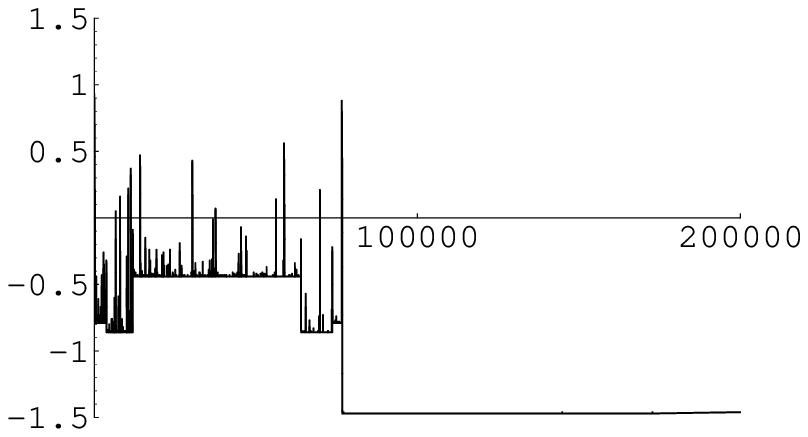, height=.25\textheight}
\end{center}
\caption{A typical random search path $(\mathbf{y}_k)_{0\leq k\leq n}$ of a L\'evy particle (l.) 
and the values of the potential function $U(\mathbf{y}_k)$ (r.). Thick lines on the left figure denote the boundaries of the
attraction domains $\Omega_i$ of the stable points $\mathbf{m}_i$, $i=1,\dots,5$.
\label{fig:1}} 
\end{figure}

We perform 100 simulations of the Markov chain \eqref{eq:yd} 
for $n=2\cdot 10^6$, 
initial conditions $\mathbf{y}_0$ distributed uniformly
in the square $[-20,20]^2$, the initial temperature $\lambda=10^{4}$, 
and the time step $h=0.1$.
The global minimum $\mathbf{m}_4$ was determined in $96$ cases ($U(\mathbf{y}_n)=-1.46$).   
The local minimum $\mathbf{m}_3$ was located twice, and each of 
the minima $\mathbf{m}_1$ and $\mathbf{m}_2$ once.
A typical random path $(\mathbf{y}_k)_{0\leq k\leq n}$ on the plain and the corresponding values of the 
energy function $U(\mathbf{y}_k)$
are shown on Figure~\ref{fig:1}.

We emphasise, that in our simulations we used only values of the potential 
$U$ without any additional information about its geometry.

\section{Conclusion and discussion}

In this paper we presented a new stochastic algorithm for global optimisation.
It allows to determine a 
global minimum of an unknown potential $U$ with help of simulated annealing of 
non-Gaussian jump-diffusions driven by the so-called stable-like processes, 
or L\'evy flights with a variable stability index $\alpha(x)$. 
We have shown that choosing $\alpha(x)$ in an appropriate way,
we can force the L\'evy particle to settle in a neighbourhood of the global maximum of $U$.
We note, that the non-constant behaviour of the stability index is crucial for the
success of the search, and a similar algorithm with usual spatially homogeneous L\'evy flights, 
i.e.\ when $\alpha(x)=\alpha_0$, leads to quite different
results, see \cite{Pavlyukevich-06b}.

Our method has the following advantages in comparison with the Gaussian 
simulated annealing considered in the introduction. 
First, the search of the global minimum is non-local,
i.e.\ when the annealed process leaves a potential well, it does not necessarily
pass to one of the neighbouring 
wells, but with strictly positive probability can jump to any
well. Moreover, the probability to jump into the deepest well is
maximal, if this well is also spatially
the biggest, which is observed in typical potential landscapes, see \cite{Schoen-97}. 
We do not expect that our algorithm would effectively detect the so-called `golf-hole'
wells. Mean transition times between the wells 
increase as a power of the large parameter $\lambda$ 
or, equivalently, the current time $t$. 
We can easily obtain theoretic estimates for a number of transitions
between the wells before settling in the deepest well. These estimates 
follow from the analysis of a
discrete time Markov chain with transition probabilities $p_{ij}=q_{ij}^{(\alpha(m_i))}[q_i^{(\alpha(m_i))}]^{-1}$, $p_{ii}=0$.

Second, we have more freedom to choose the parameters of the system. 
Indeed, if the values of $U(m_i)$ are not known, there is no method 
which helps to determine the cooling rate
$\theta$. (One has the same problem to determine $\hat\theta$ in 
\eqref{eq:hattheta} in Gaussian case.) 
However, in our algorithm, $\theta$ is 
chosen together with a variable stability index
$\alpha(x)$.

Third, our method allows to drive the L\'evy particle into any well whose
location is approximately known. 
We can also determine a local minimum with energy below the certain given value.
The choice of parameters in these regimes is independent of 
the geometry of the potential.
Such search regimes are not possible in the classical setting.

Finally, the temperature decreases polynomially fast in time, i.e.\
$\sim t^{-\theta}$, and not logarithmic. This significantly increases the 
accuracy of empirical estimates for the local minima locations.

Although the theoretic basis for the success of the search is established,
many questions are still open. 
For example, we have to understand how to choose the optimal pair $\alpha(x)$ and
$\theta$, which minimises the search time. 
Indeed, if $m_i$ is not a global minimum, we can reduce the life time of the 
particle in the neighbourhood of $m_i$ making $\alpha(m_i)$ small.
On the other hand, in this case, the process $Y$
will tend to make very big jumps, and thus jump out to one of the peripheral wells. 
As a consequence, the search would be slow, if the global minimum of $U$ is
attained in one of the inner wells. Thus, the value of $\alpha(m_i)$ should not
be very small to exclude very big jumps, and should be well separated from
$\alpha(M)$ to block trapping in the false well.
The problem of very big jumps can be also avoided by consideration of truncated
L\'evy flights with maximal jump size not exceeding the size of the search domain.
However, in this case the simulation of random input can be more complicated.
We shall address these and other questions in our further research. 

\medskip

\textbf{Acknowledgements.} This work was supported by the DFG research project
\textit{Stochastic Dynamics of Climate States}. The author thanks P.\ Imkeller
for stimulating discussions.


\bibliography{biblio-new}
\bibliographystyle{alpha}

\bigskip

\parbox{.48\linewidth}{
\noindent
\textsc{Ilya Pavlyukevich}\\
Department of Mathematics\\
Humboldt University of Berlin\\
Rudower Chaussee 25\\
12489 Berlin Germany\\
E.mail: \texttt{pavljuke@mathematik.hu-berlin.de}\\
http://www.mathematik.hu-berlin.de/\~{}pavljuke
}\hfill
\parbox{.5\linewidth}{
\noindent
}

\end{document}